\documentstyle[12pt]{article}
\newcommand{\be}{\begin{equation}}
\newcommand{\ee}{\end{equation}}
\newcommand{\bdm}{\begin{displaymath}}
\newcommand{\edm}{\end{displaymath}}
\newcommand{\bea}{\begin{eqnarray}}
\newcommand{\eea}{\end{eqnarray}}

\newcommand{\fs}{\; \; .}
\newcommand{\co}{\; \; ,}

\newcommand{\eff}{{e\hspace{-0.1em}f\hspace{-0.18em}f}}

\newcommand{\QCD}{\mbox{\scriptsize Q\hspace{-0.1em}CD}}
\newcommand{\indR}{\mbox{\scriptsize R}}
\newcommand{\indL}{\mbox{\scriptsize L}}

\newcommand{\lvac}{\langle 0|\,}
\newcommand{\rvac}{\,|0\rangle}

\newcommand{\Hbar}{\,\overline{\rule[0.7em]{0.6em}{0em}}\hspace{-0.8em}H}
\newcommand{\ubar}{\overline{\rule[0.42em]{0.4em}{0em}}\hspace{-0.5em}u}
\newcommand{\dbar}{\,\overline{\rule[0.65em]{0.4em}{0em}}\hspace{-0.6em}d}
\newcommand{\sbar}{\overline{\rule[0.42em]{0.4em}{0em}}\hspace{-0.5em}s}
\newcommand{\cbar}{\overline{\rule[0.42em]{0.4em}{0em}}\hspace{-0.5em}c}

\newcommand{\Qvec}{\vec{\hspace{0.04em}Q}}
\newcommand{\Mvec}{\vec{\hspace{0.04em}M}}
\def\im#1{{\mbox{\scriptsize #1}}}

\begin{document}
\begin{titlepage}
\begin{flushright}BUTP-94/17\end{flushright}
\rule{0em}{2em}\vspace{2em}
\begin{center}
{\LARGE {\bf Goldstone Bosons}}\\ \vspace{2em}
H. Leutwyler\\Institut f\"{u}r theoretische Physik der Universit\"{a}t
Bern\\Sidlerstr. 5, CH-3012 Bern, Switzerland\\
\vspace{2em}
September 1994\\
\vspace{3em}
{\bf Abstract} \\
\vspace{2em}
\parbox{30em}{
The paper concerns the effective
field theory methods used to study the low energy structure of systems with a
spontaneously broken symmetry. I first explain how the method works
in the context of quantum chromodynamics and then discuss a few
general aspects, related to the universality of effective theories. In
particular, I compare some of the effective field theories used in
condensed matter physics with those relevant for particle physics.}\\

\vspace{10em}
\rule{30em}{.02em}\\
{\footnotesize Work
supported in part by Schweizerischer Nationalfonds}
\end{center}
\end{titlepage}

Goldstone bosons occur in many areas of physics. I first discuss the
phenomenon in the context of the strong interaction, where the
pions play the role of the Goldstone bosons. Later
on, I will identify those features which are independent of the particular
system under consideration and compare the
effective field theory used to analyze the low energy properties of
the strong interaction with some of the effective theories
encountered in condensed matter physics.

The strong
interaction is mediated by the gauge field of colour, which binds the quarks
to colour neutral bound states such as the proton or the neutron. The structure
of the relevant gauge field theory, quantum chromodynamics, is similar to
the one describing the electromagnetic interaction.
As it is the case with the coupling between photons and electrons,
the interaction of the gluons with the quarks is fully
determined
by gauge invariance. This implies, in particular, that the various different
quark flavours, $u,d,\ldots$ interact with the gluons in precisely the same
manner. As far as the strong interaction is
concerned, the only distinction
between, say, an $s$-quark and a $c$-quark is that the mass is different.
In this respect, the situation is the same as in
electrodynamics, where the interaction of the charged leptons with the
photon is also universal, such that the only difference between
$e,\mu$ and $\tau$ is the mass.
As an immediate consequence, the properties of a bound state like the
$\Lambda_s=(uds)$ are identical with those of the $\Lambda_c=(udc)$,
except for the fact that $m_c$ is larger than $m_s$.
\section{Isospin symmetry}
\label{iso}
A striking property of the observed pattern of bound states is that they
come in nearly degenerate {\it isospin} multiplets: $(p,n),$
$(\pi^+,\pi^0,\pi^-),\,(K^+,K^0),
\;\ldots\;$
In fact, the splittings within these multiplets are so small that, for a long
time, isospin was taken for an
{\it exact} symmetry of the strong interaction; the observed
small
mass difference between neutron and proton or $K^0$ and $K^+$ was blamed on the
electromagnetic interaction.
We now know that this picture is incorrect: the
bulk of isospin breaking does not originate in the electromagnetic
fields, which surround the various particles, but is due to the fact that
the $d$-quark is somewhat heavier than the $u$-quark.

{}From a theoretical point of view, the quark masses are free
parameters ---
QCD makes sense for any value of $m_u,m_d,\ldots\;$ It
is perfectly
legitimate to compare the real situation with a theoretical one, where some of
the quark masses are given values, which differ from those found in
nature. In connection with isospin symmetry, the theoretical limiting case
of interest is a fictitious world, with
$m_u=m_d$. In this limit, the flavours
$u$ and $d$ become indistinguishable. The
Hamiltonian acquires an exact symmetry with respect to the transformation
\bdm \begin{array}{lll}u&\rightarrow&\alpha u +\beta d\\
d&\rightarrow&\gamma u + \delta d\end{array}\co\hspace{5em}
V=\left(\!\begin{array}{ll}\alpha&\!\!\beta\\
\gamma&\!\!\delta\end{array}\!\right)\co\edm
provided the $2\!\times\! 2$ matrix $V$ is unitary, $V\!\in\!\mbox{U(2)}$. Even
for $m_u\!\neq\! m_d$, the Hamiltonian of QCD is invariant under
a change of phase of the quark fields. The extra
symmetry, occurring if the masses of $u$ and $d$ are taken to be the same, is
contained in the subgroup SU(2), which results if the phase of the
matrix $V$ is subject to
the condition $\det V\!=\!1$. The above transformation law states that $u$ and
$d$ form an isospin doublet, while the remaining
flavours $s,\,c,\,\ldots\,$ are singlets.

In reality, $m_u$ differs from $m_d$. The isospin group SU(2) only represents
an {\it approximate} symmetry. The piece
of the QCD Hamiltonian, which breaks isospin symmetry, may be exhibited by
rewriting the mass term of the $u$ and $d$ quarks in the form
\bdm m_u\,\ubar u+m_d\,\dbar d=\mbox{$\frac{1}{2}$}(m_u+m_d)(\ubar u+
\dbar d)+\mbox{$\frac{1}{2}$}(m_d-m_u)(\dbar d-
\ubar u)\fs\edm
The remainder of the Hamiltonian is invariant
under isospin transformations and the same is true of the operator
$\ubar u+\dbar d$. The
QCD Hamiltonian thus consists of an isospin invariant part
$\Hbar_0$ and a
symmetry breaking term $\Hbar_{\mbox{\scriptsize sb}}$, proportional to
the mass difference $m_d-m_u$, \be \label{iso1}
H_{\QCD}=\,\Hbar_0
+\,\Hbar_{\mbox{\scriptsize sb}}\co\hspace{3em}
\Hbar_{\mbox{\scriptsize sb}}\,=\mbox{$\frac{1}{2}$}(m_d - m_u)
\int\!\!d^3\!x\, (\dbar d - \ubar u)\fs\ee
The strength of isospin breaking is controlled by the quantity
$m_d-m_u$, which plays the role of a {\it symmetry breaking
parameter}. The fact that the multiplets are nearly degenerate
implies that the operator $\Hbar_{\mbox{\scriptsize sb}}$
only represents a small
perturbation --- the mass difference $m_d-m_u$ must be very small.
QCD thus provides a remarkably simple explanation for the fact that the strong
interaction is nearly invariant under
isospin rotations: it so happens that the difference between $m_u$ and
$m_d$
is small and this is all there is to it.

The symmetry breaking also shows up in the properties of the vector currents,
e.g. in those of $\ubar \gamma^\mu d$. The integral of the corresponding charge
density over space, $I^+=\int\!d^3\!x\,u^\dagger d$, is the isospin raising
operator, converting a $d$-quark into a $u$-quark. The divergence of the
current is given by
\be\label{iso2} \partial_\mu (\ubar \gamma^\mu d)
=i\,(m_u-m_d)\,\ubar d\co \ee
and only vanishes for $m_u=m_d$, the condition for the charge $I^+$ to
be conserved. In the symmetry limit, there are three such conserved charges,
the three components of isospin,
$\vec{\rule{0em}{0.65em}}\hspace{-0.25em}I=(I^1,I^2,I^3)$. The
isospin raising operator considered above is the combination $I^+=I^1+i\,I^2$.
Since $\Hbar_0$ is invariant under isospin rotations, it conserves isospin,
\be\label{iso3}
[\;\,\vec{\rule{0em}{0.65em}}\hspace{-0.25em}I\,,\,\Hbar_0\,]=0\fs\ee
\section{Chiral symmetry}
\label{chir}
The approximate symmetry of the Hamiltonian explains
why the bound states of QCD exhibit a multiplet pattern,
but does not account
for an observation, which is equally striking and which plays a crucial role in
strong interaction physics --- the mass gap of the theory, $M_\pi,\,$ is
remarkably small.
The approximate symmetry, hiding behind this observation, was discovered by
Nambu \cite{Nambu}. It originates in a phenomenon, which is well-known
from neutrino physics: right- and left-handed components of {\it
massless} fermions do not communicate.

The symmetry, which forbids right-left-transitions,
manifests itself in the properties of the axial vector currents, such as
$\ubar \gamma^\mu\gamma_5d$. The corresponding continuity equation reads
\be\label{chir1}  \partial_\mu(\ubar \gamma^\mu\gamma_5d)=i\,(m_u+m_d)\,
\ubar\gamma_5d\fs\ee
While the divergence of the vector current
$\ubar \gamma^\mu d$ is proportional to the difference $m_u-m_d$, the
one of the axial current is proportional to
the sum $m_u+m_d$. If the two masses are
set equal, the vector current is conserved and the
Hamitonian becomes symmetric with respect to isospin rotations. If they are
not only taken equal, but equal to zero, then the axial current is
conserved, too, such that the corresponding charge
$I_5^+=\int\!d^3\!x\,d^\dagger\gamma_5 u$ also commutes with the
Hamiltonian --- QCD acquires an additional symmetry.

The isospin operator $I^+$ converts a $d$-quark
into a $u$-quark, irrespective of the helicity. The operator $I_5^+$,
however, acts differently on
the right- and left-handed components. The sum $\frac{1}{2}(I^++I_5^+)$
takes a righthanded $d$-quark into a righthanded $u$-quark, but leaves
left-handed ones alone.
This implies that, for massless quarks, the
Hamiltonian is invariant with
respect to {\it chiral}
transformations: independent
isospin rotations of the right- and left-handed components of $u$ and $d$,
\bdm\left(\!\!\begin{array}{c} u_{\indR}\\d_{\indR}
\end{array}\!\!\right)\rightarrow V_{\indR}
\left(\!\!\begin{array}{c} u_{\indR}\\d_{\indR}
\end{array}\!\!\right)\co\;\;\;\;
\left(\!\!\begin{array}{c} u_{\indL}\\d_{\indL}
\end{array}\!\!\right)\rightarrow V_{\indL}
\left(\!\!\begin{array}{c} u_{\indL}\\d_{\indL}
\end{array}\!\!\right)\co\;\;\;\;\;\;\;V_{\indR},\,
V_{\indL} \in \mbox{SU(2)}\fs\edm
The corresponding symmetry group is the direct product of two separate isospin
groups, SU(2)$_{\indR}\times$SU(2)$_{\indL}$.
The symmetry is
generated two sets of isospin operators: ordinary isospin,
$\,\vec{\rule{0em}{0.65em}}\hspace{-0.25em}I$ and
chiral isospin, $\,\vec{\rule{0em}{0.65em}}\hspace{-0.25em}I_5$. The
particular operator considered above is
the linear combination $I_5^+=I^{\,1}_5+i\,I^{\,2}_5$.

In reality,
chiral symmetry is broken, because $m_u$ and $m_d$ do not vanish.
As above, the Hamiltonian may be split into a
piece which is invariant under the symmetry group of interest and a piece
which breaks the symmetry. In the present case, the symmetry breaking part
is the full mass term of the $u$ and $d$ quarks,
\be\label{chir2} H_{\QCD}=H_0+
H_{\mbox{\scriptsize sb}}\;\;,\;\;
H_{\mbox{\scriptsize sb}}
=\int\!d^3\!x(m_u\,\ubar u+m_d\,\dbar d)\fs\ee
The symmetric part conserves ordinary as well as chiral isospin,
\be\label{chir3} [\;\,\vec{\rule{0em}{0.65em}}\hspace{-0.25em}I \,
,\,H_0\,]=0\co\hspace{3em}
 [\;\,\vec{\rule{0em}{0.65em}}\hspace{-0.25em}I_5
,\,H_0\,]=0\fs\ee
Note that the symmetry group exclusively
acts on $u$ and $d$ --- the
remaining quarks $s,c,\ldots$ are singlets. The corresponding mass terms
$m_s\sbar s+m_c\,\cbar c+\ldots$ do not break the symmetry and are included in
$H_0$.
\section{Spontaneous symmetry breakdown}
\label{sp}
Much before QCD was discovered, Nambu pointed out that chiral symmetry
breaks down spontaneously.
The phenomenon plays a crucial role for the properties
of the strong interaction at low energy. To discuss it, I return to
the theoretical scenario,
where $m_u$ and $m_d$ are set equal to zero.

In this framework, isospin is conserved. The isospin group SU(2)
represents the prototype of a "manifest"
symmetry, with all the consequences known from quantum mechanics: (i) The
energy
levels form degenerate multiplets. (ii) The operators
$\,\vec{\rule{0em}{0.65em}}\hspace{-0.25em}I$ generate transitions within the
multiplets, taking a neutron, e.g., into a proton,
$I^+|n\rangle=|p\rangle$. (iii) The ground state is an isospin
singlet,
\be\label{sp1} \vec{\rule{0em}{0.65em}}\hspace{-0.25em}I\rvac=0\fs\ee

If chiral symmetry was realized in the same manner, the energy levels would
occur in degenerate multiplets of
the group SU(2)$_{\indR}\times$SU(2)$_{\indL}$. Since the chiral isospin
operators $\,\vec{\rule{0em}{0.65em}}\hspace{-0.25em}I_5$
carry negative parity, the multiplets would then necessarily contain members of
opposite parity. The listings of the Particle Data Group, however, do not show
any trace of such a pattern. A particle with the quantum
numbers of $I_5^+|n\rangle$ and nearly the same mass as the neutron, e.g.,
is not observed in nature.

In fact, the symmetry of the Hamiltonian does not ensure that the
corresponding
eigenstates form multiplets of the symmetry group. In particular, the
state with the lowest eigenvalue of the Hamiltonian need not be a
singlet. In the case of a magnet, e.g., the Hamiltonian is
invariant under rotations of the spin directions, but the ground state fails
to be invariant, because
the spins are aligned and thereby single out a direction. Whenever the state
with the lowest eigenvalue is less symmetric than the Hamiltonian, the symmetry
is called "spontaneously broken" or "hidden". Chiral symmetry belongs to
this category. For dynamical reasons, the most important state --- the
vacuum --- is symmetric only under ordinary isospin rotations, but does not
remain invariant if a chiral rotation is applied,
\be\label{sp2}\vec{\rule{0em}{0.65em}}\hspace{-0.25em}I_5\rvac\neq 0\fs\ee
Since the Hamiltonian commutes with chiral isospin, the three
states
$\,\vec{\rule{0em}{0.65em}}\hspace{-0.25em}I_5\rvac$ have the same energy
as the vacuum, $E=0$. The operators
$\,\vec{\rule{0em}{0.65em}}\hspace{-0.25em}I_5$ do
not carry momentum, either, so that the states
$\,\vec{\rule{0em}{0.65em}}\hspace{-0.25em}I_5\rvac$ have $\vec{P}=0$.
This indicates that the
spectrum of physical states contains three
massless particles. Indeed, the Goldstone theorem \cite{Goldstone}
rigorously shows that spontaneous symmetry breakdown gives rise to
massless particles, "Goldstone
bosons". Their quantum numbers are those of the states
$\rule{0.3em}{0em}\vec{\rule{0em}{0.65em}}\hspace{-0.25em}I_5\rvac$: spin
zero, negative parity and $I=1$.

The three lightest mesons, $\pi^+\!,\pi^0\!,\pi^-$, carry precisely these
quantum numbers. The chiral isospin operators act like creation or
annihilation operators for pions: Applied to the vacuum, they generate a
state containing a pion, $I_5^+\rvac=|\pi^+\rangle$. Applied to a
neutron, they do not lead to a parity partner, but instead yield a state
containing a neutron and a pion, $I_5^+|n\rangle=|n\pi^+\rangle$, etc.
\section{Pion mass}
\label{pm}
The above discussion concerns the theoretical world, where $u$ and $d$ are
assumed to be massless, such that the group
SU(2)$_{\indR}\times$SU(2)$_{\indL}$ represents an exact symmetry.
The Hamiltonian of QCD contains a quark mass term,
which breaks the symmetry.
To see how this affects the mass of the
Goldstone bosons,
consider the transition matrix element of the axial current
$\ubar\gamma^\mu\gamma_5d$,
from the vacuum to a one-pion state. Lorentz invariance
implies that this matrix element is determined by the pion
momentum $p^\mu$, up to a constant,
\bdm\langle\pi^+(p)|\,\ubar(x)\gamma^\mu\gamma_5d(x)\rvac
=-ip^\mu\sqrt{2}\,F_\pi\, e^{ipx}\fs\edm
The value of the constant is measured in pion decay, $F_\pi\simeq
93\;\mbox{MeV}$. For the divergence
$\partial_\mu(\ubar\gamma^\mu\gamma_5d)$, this yields an expression
proportional to
$p^2=M_{\hspace{-0.07em}\pi^{\hspace{-0.07em}+}}^2$. Denoting the analogous
matrix element of the pseudoscalar density by $G_\pi$,
\bdm\langle\pi^+(p)|\,\ubar(x)\gamma_5d(x)\rvac
=\; i\sqrt{2}\,G_\pi\, e^{ipx}\co\edm
the conservation law (\ref{chir1}) thus implies the exact
relation
\be\label{pm1}
M_{\hspace{-0.07em}\pi^{\hspace{-0.07em}+}}^2=(m_u+m_d)\,(G_\pi/ F_\pi)\fs\ee
The relation confirms that, when the symmetry breaking parameters $m_u,m_d$
are put equal to zero, the pion mass vanishes, independently of the
masses of the other quark flavours.
The group SU(2)$_{\indR}\times$SU(2)$_{\indL}$ then represents a
spontaneously broken, {\em exact} symmetry, with three strictly massless
Goldstone bosons. When the quark
masses are turned on, the Goldstone bosons pick up mass:
$M_{\hspace{-0.07em}\pi^{\hspace{-0.07em}+}}$ grows in
proportion to
\raisebox{0.25em}{$\sqrt{\rule{3.8em}{0em}}$}$\hspace{-3.8em}m_u+m_d\,$.
The pions remain light, provided $m_u$ and
$m_d$ are small. The quark mass term of the Hamiltonian then
amounts to a small perturbation, such that the
group SU(2)$_{\indR}\times$SU(2)$_{\indL}$ still represents
an {\em approximate} symmetry, with
approximately massless
Goldstone bosons.

The decomposition of the QCD Hamiltonian in eq. (\ref{chir2}) may
be compared with the standard perturbative splitting
\bdm H_{\QCD}=H_{\mbox{\scriptsize free}}+ H_{\mbox{\scriptsize int}}\co\edm
where the first term describes free quarks and gluons, while the second
accounts for their interaction. The corresponding expansion parameter is
the coupling constant $g$. Since QCD
is asymptotically free, the effective coupling becomes weak
at large momentum transfers --- processes which exclusively involve large
momenta may indeed be analyzed by treating the interaction as a
perturbation. Perturbation theory, however, fails in
the low energy domain,
where the effective coupling is strong, such that it is not meaningful to
truncate the expansion in powers of $H_{\mbox{\scriptsize int}}$ after the
first few
terms. In particular, the structure of the ground state cannot be analyzed in
this way, while the above decomposition, which retains the
interaction
among the quarks and gluons in the "unperturbed" Hamiltonian $H_0$ and
only treats $m_u$ and $m_d$
as perturbations, is perfectly suitable for that purpose.
Note that the character of the
perturbation series
in powers of $H_{\mbox{\scriptsize sb}}$ is quite different from the
one in powers of $H_{\mbox{\scriptsize int}}$: while the eigenstates of
$H_{\mbox{\scriptsize free}}$ are
known explicitly, this is not the case with $H_0$, which still describes
a highly nontrivial, interacting system. $H_0$ differs from the full
Hamiltonian only in one respect: it possesses an exact group of chiral
symmetries.
\section{Effective field theory}
\label{eff}
At low energies, the behaviour of scattering amplitudes or current matrix
elements can be described in terms of a {\it Taylor series expansion} in powers
of the momenta.
The electromagnetic form factor of the pion, e.g., may be
exanded in powers of the momentum transfer $t$.
In this case, the first two Taylor coefficients are related to the total charge
of the particle and to the mean square radius of the charge distribution,
respectively,
\be \label{taylor}
f_{\pi^+}(t) = 1 + \mbox{$\frac{1}{6}$} \langle r^2\rangle_{\pi^+}\, t +
O(t^2)\fs \end{equation}
Scattering lengths and effective ranges are analogous low energy
constants occurring in the Taylor series expansion of scattering amplitudes.

The occurrence of light particles gives rise to singularities in the low
energy domain, which limit the range of validity of the Taylor series
representation. The form factor $f_{\pi^+}(t)$, e.g., contains a branch cut
at $t=4 M_\pi^2$, such that the formula (\ref{taylor}) provides an adequate
representation only for $|\,t\,|\ll 4 M_\pi^2$. The problem becomes even more
acute if $m_u$ and $m_d$ are set equal to zero. The pion mass then
disappears, the branch cut sits at $t=0$ and the Taylor series does not
work at all. I first discuss the method used in the low energy analysis for
this extreme case, returning
to the physical situation with $m_u,m_d\neq 0$ below.

The reason why the spectrum of QCD with two massless quarks contains three
massless bound states is understood:
they are the Goldstone bosons of a
hidden symmetry. The symmetry, which
gives birth to these, at the same time also determines their low energy
properties. This makes it possible to explicitly work out
the poles and branch cuts generated by the exchange of Goldstone bosons.
The remaining singularities are
located comparatively far from the origin, the nearest one being due to
the $\rho$-meson. The result is a modified Taylor series expansion in powers
of the momenta, which works, despite the presence of massless particles.
In the case of the $\pi\pi$ scattering amplitude,
e.g., the radius of convergence of the modified series
is given by $s=M_\rho^2$, where $s$ is the square of the energy in the center
of mass system (the first few
terms of the series only yield a decent description of the amplitude if
$s$ is smaller than the radius of
convergence, say $s\!<\!\frac{1}{2}M_\rho^2\rightarrow \sqrt{s}\!<
540\;\mbox{MeV}$).

As pointed out by Weinberg \cite{Weinberg79}, the modified expansion
may explicitly be constructed by means of an effective field theory, which is
referred to as {\it chiral perturbation theory} and involves the following
ingredients: \\ (i) The quark and gluon fields of QCD are
replaced by a set of pion fields, describing the degrees of freedom of the
Goldstone
bosons. It is convenient to collect these in a
$2\!\times\!2$
matrix U$(x)\!\in\,$SU(2). \\
(ii) The Lagrangian of QCD is replaced by an
effective Lagrangian, which only involves the field U$(x)$, and its derivatives
\bdm {\cal L}_{\QCD}\;\;\longrightarrow \;\;{\cal L}_\eff(U,\partial
U,\partial^2U,\ldots)\fs\edm
(iii) The low energy expansion corresponds to an
expansion of the effective
Lagrangian, ordered according to the number of derivatives of the field
$U(x)$.
Lorentz invariance only permits terms with an even number
of derivatives,
\bdm
{\cal L}_{\eff}= {\cal L}_{\eff}^{\,2} +
{\cal L}_{\eff}^{\,4} + {\cal L}_{\eff}^{\,6} +
\ldots
\edm

Chiral symmetry very strongly constrains the form of the terms occurring
in the series. In particular, it excludes momentum
independent interaction vertices:
Goldstone bosons can only interact if they carry momentum. This property
is essential for the consistency of the low energy analysis, which treats the
momenta as expansion parameters.
The leading contribution involves two derivatives,
\be \label{eff1}
{\cal L}_{\eff}^{\,2} = \mbox{$\frac{1}{4}$}F_\pi^2 \mbox{tr} \{
\partial_\mu U^+ \partial^\mu U \} \co
\ee
and is fully determined by the pion decay constant. At order $p^4$, the
symmetry permits two independent terms,\footnote{In the framework of the
effective theory, the anomalies of QCD
manifest themselves through an extra contribution,
the Wess-Zumino term, which is also of order $p^4$ and is proportional to the
number of colours.}
\be\label{eff3} {\cal L}_{\eff}^{\,4}=\mbox{$\frac{1}{4}$}l_1 (\mbox{tr} \{
\partial_\mu U^+ \partial^\mu U \})^2
+ \mbox{$\frac{1}{4}$}l_2\mbox{tr} \{
\partial_\mu U^+ \partial_\nu U \}\mbox{tr} \{
\partial^\mu U^+ \partial^\nu U \}\co\ee
etc. For most applications, the derivative expansion
is needed only to this order.

The most remarkable property of the method is that it does not
mutilate the theory under investigation:
The effective field theory framework is no more
than an efficient machinery, which
allows one to work out the modified Taylor series, referred to above.
If the effective Lagrangian includes all of the terms
permitted by the symmetry, the
effective theory is mathematically equivalent to QCD \cite{Weinberg79,found}.
It exclusively exploits the symmetry properties of QCD and involves an infinite
number of effective coupling constants,
$F_\pi,l_1,l_2,\ldots\;$, which represent the Taylor coefficients of the
modified expansion.

In QCD, the
symmetry, which controls the low energy properties of the Goldstone bosons, is
only an approximate one. The constraints imposed by the hidden,
approximate symmetry can still be worked out, at the price of
expanding the
quantities of physical interest in powers of the symmetry breaking parameters
$m_u$ and $m_d$. The low energy analysis then involves a combined
expansion,
which treats both, the momenta and the quark masses as small parameters.
The effective Lagrangian picks up additional terms, proportional to powers of
the quark mass matrix,
\bdm
m = \left(\mbox{\raisebox{0.4em}{$ m_u $}}\;\mbox{\raisebox{-0.4em}
{$m_d$}}\,
\right)
\edm
It is convenient to count $m$ like two powers of
momentum, such that the expansion of the effective Lagrangian still starts at
$O(p^2)$ and only contains even
terms. The leading contribution picks up a term linear in $m$,
\be\label{eff2}
{\cal L}_{\eff}^{\,2} = \mbox{$\frac{1}{4}$}F_\pi^2 \mbox{tr} \{
\partial_\mu U^+ \partial^\mu U \} +\mbox{$\frac{1}{2}$}F_\pi^2B\,\mbox{tr}
\{m(U+U^\dagger)\}\fs
\ee
Likewise, ${\cal L}_\eff^4$ receives additional contributions, involving two
further effective coupling constants, $l_3,l_4$, etc.

The expression (\ref{eff2}) represents a compact summary of the soft pion
theorems
established in the 1960's: The leading terms in the low energy expansion of the
scattering amplitudes and current matrix elements are given by the tree graphs
of this Lagrangian.
The coupling constant $B$ is needed to account for the symmetry
breaking effects generated by the quark masses at leading order. It represents
the coefficient of the leading term in the expansion of
the pion mass in powers of $m_u$ and $m_d$, $M_\pi^2=(m_u+m_d)B+O(m^2)$.
According
to section 4, the same constant also determines the vacuum-to-pion matrix
element of
the pseudoscalar density, $G_\pi =F_\pi B +O(m)$. Furthermore, the relation of
Gell-Mann, Oakes and Renner, $F_\pi^2M_\pi^2=-(m_u+m_d)\,\lvac\ubar u\rvac
+O(m^2)$,
which immediately follows from the
above expression for the effective Lagrangian, shows that the magnitude of
the quark condensate is also related to the value of $B$.

The effective field theory
represents an efficient and systematic framework, which allows one to work out
the corrections to the soft pion predictions, those arising from the
quark masses as well as those from the terms of higher order
in the momenta. The evaluation is based on a perturbative
expansion of the quantum fluctuations of the effective field. In addition to
the tree graphs relevant for the soft pion results, graphs containing vertices
from the higher order contributions ${\cal L}_\eff^4,{\cal L}_\eff^6\ldots$ and
loop graphs
contribute. The leading term of the effective Lagrangian describes
a nonrenormalizable theory, the "nonlinear $\sigma$-model". The
higher order terms in the derivative expansion, however, automatically contain
the relevant counter terms. The divergences occurring in the loop
graphs merely renormalize the effective coupling constants. The effective
theory is a perfectly renormalizable scheme, order by order in the low
energy expansion and the results obtained with it are
independent of the regularization used.
\section{Universality}
\label{uni}
The properties of the effective theory are governed by the hidden symmetry,
which is responsible for the occurrence of Goldstone bosons. In
particular, the form of the effective Lagrangian only depends on the symmetry
group G of the Hamiltonian and on the subgroup $\mbox{H}\subset \mbox{G}$,
under which the ground state is invariant. The Goldstone bosons live on the
difference between the two
groups, i.e., on the quotient G/H. The specific dynamical properties of the
underlying theory do not play any role. To discuss the consequences of this
observation,
I again assume that G is an exact symmetry.

In the case of QCD with two
massless quarks, $\mbox{G}=\mbox{SU(2)}_{\indR}\times\mbox{SU(2)}_{\indL}$ is
the group of
chiral isospin rotations, while $\mbox{H}=\mbox{SU(2)}$ is the ordinary isospin
group.
The Higgs model is another example of a theory with spontaneously broken
symmetry. It plays a crucial role in the Standard Model, where it describes
the generation of mass. The model involves a scalar field
$\vec{\phi}$ with four components. The Hamiltonian is invariant under
rotations of the vector $\vec{\phi}$, which form the group G = O(4). Since the
field picks up a vacuum expectation value, the
symmetry is spontaneously broken to the subgroup of those rotations,
which leave the vector $\lvac\vec{\phi}\rvac$ alone, H = O(3).
It so happens that these groups are the same as those above,
relevant for QCD.\footnote{The structure of the effective Lagrangian rigorously
follows from the
Ward identities for the Green functions of the currents, which also reveal the
occurrence of anomalies \cite{found}. The
form of the Ward identities is controlled by
the structure of G and H in the infinitesimal neighbourhood of the
neutral element. In this sense, the symmetry groups of the two models are the
same: O(4) and O(3) are {\it locally}
isomorphic to SU(2)$\times$SU(2) and
SU(2), respectively. }
The fact that the symmetries are the same implies that
the effective field theories are identical: (i) In either
case, there are three
Goldstone bosons, described by a matrix field $U(x)\in\mbox{SU(2)}$. (ii) The
form of the effective Lagrangian is precisely the same.
In particular, the expression
\bdm
{\cal L}_{\eff}^{\,2} = \mbox{$\frac{1}{4}$}F_\pi^2 \mbox{tr} \{
\partial_\mu U^+ \partial^\mu U \}
\edm
is valid in either case. At the level of the effective theory, the
only
difference between these two physically quite distinct models is that
the numerical values of the effective coupling constants are different.
In the case of QCD, the one occurring at leading order of the
derivative expansion is the pion decay constant, $F_\pi\simeq
93\,\mbox{MeV}$, while in the Higgs model, this coupling constant is larger
by more than three orders of magnitude, $F_\pi\simeq
250\;\mbox{GeV}$. At next-to-leading order, the effective coupling constants
are also different; in particular, in QCD, the anomaly coefficient is equal to
$\mbox{N}_c$, while in the Higgs model, it vanishes.

As an illustration, I compare the condensates of the two theories, which
play a role
analogous to the spontaneous magnetization $\langle\Mvec\rangle$ of a
ferromagnet (or the staggered magnetization of an antiferromagnet).
At low temperatures, the magnetization singles out a direction --- the ground
state spontaneously breaks the symmetry
of the Hamiltonian with respect to rotations. As the system is heated, the
spontaneous magnetization decreases, because the thermal disorder acts against
the alignment of the spins. If the temperature is high enough, disorder
wins, the spontaneous magnetization disappears and rotational symmetry is
restored. The temperature at which this happens is the Curie temperature.
Quantities, which allow one to distinguish the ordered from the disordered
phase are called {\it order parameters}. The magnetization is the prototype of
such a parameter.

In QCD, the most important order parameter (the one of lowest dimension) is the
quark condensate. At nonzero temperatures, the condensate is given
by the thermal expectation
value \bdm \langle\ubar u\rangle_{\hspace{-0.05em}\mbox{\raisebox{-0.2em}
{\scriptsize $T$}}} =\frac{\mbox{Tr}\{\,\ubar u
\exp (-\,H/kT)\}}{ \mbox{Tr}\{\exp(-\,H/kT)\} }\fs
\edm
The condensate melts if the temperature is
increased. At a critical temperature, somewhere in the range
$140\,\mbox{MeV}\!<\!T_c\!<\!\mbox{180}\;\mbox{MeV}$, the quark condensate
disappears and chiral symmetry is restored. The same qualitative
behaviour also occurs in the Higgs model, where the expectation value
$\langle\,\vec{\phi}\,
\rangle_{\hspace{-0.05em}\mbox{\raisebox{-0.2em}
{\scriptsize $T$}}}$ of the scalar field represents the most prominent order
parameter.

At low temperatures, the thermal trace is dominated by
states of low energy. Massless particles generate contributions which are
proportional to powers of the temperature, while massive ones like the
$\rho$-meson are suppressed by the corresponding Boltzmann factor,
$\exp(-M_\rho/kT)$. In the case of a spontaneously broken symmetry,
the massless particles are the Goldstone bosons and their contributions may be
worked out by means of effective field theory. For the quark condensate, the
calculation has been done \cite{Gerber}, up to and including terms of order
$T^6$: \be\label{cond1}\langle\ubar
u\rangle_{\hspace{-0.05em}\mbox{\raisebox{-0.1em} {\scriptsize $T$}}} =
\lvac\ubar u\rvac\!
\left\{1\,-\,\frac{T^2}
{8F_\pi^2 }
\,-\,\frac{T^4}{384F_\pi^4}
\,-\,\frac{T^6}{288F_\pi^6}
\, \ln(T_1/T)
\,+\,O(T^8)\right\}\fs\ee
The formula is exact --- for massless quarks, the temperature scale relevant
at low $T$ is the pion decay constant. The additional logarithmic scale $T_1$
occurring at order $T^6$ is determined by the effective coupling constants
$l_1,l_2$, which enter the expression (\ref{eff3}) for the effective Lagrangian
of order $p^4$. Since these are known from the phenomenology of $\pi\pi$
scattering, the value of $T_1$ is also known:
$T_1=470\pm110\;\mbox{MeV}$.

Now comes the point I wish to make. The effective Lagrangians
relevant for QCD and for the Higgs model are the same. Since the
operators of which we are considering the expectation values also transform in
the
same manner, their low temperature expansions are identical. The above formula
thus holds, without any change whatsoever, also for the Higgs condensate,
\bdm \langle\,\vec{\phi}\,\rangle_{\hspace{-0.1em}\mbox{\raisebox{-0.2em}
{\scriptsize $T$}}} =
\lvac\vec{\phi}\rvac\!
\left\{1\,-\,\frac{T^2}
{8F_\pi^2 }
\,-\,\frac{T^4}{384F_\pi^4}
\,-\,\frac{T^6}{288F_\pi^6}
\, \ln(T_1/T)
\,+\,O(T^8)\right\}\fs\edm
In fact, the universal term of order $T^2$ was discovered in the framework of
this model, in connection with work on the electroweak phase transition
\cite{Binetruy}.

These examples illustrate the physical nature of effective theories: At long
wavelength, the microscopic structure does not play any role. The behaviour
only depends on those degrees of freedom, which require little
excitation
energy. The hidden symmetry, which is responsible for the absence of an
energy gap and for the occurrence of Goldstone bosons, at the same time also
determines their low energy properties. For this reason, the form of
the effective Lagrangian is controlled
by the symmetries of the system and is, therefore, universal.
The microscopic structure of the underlying theory exclusively manifests itself
in the numerical values of the effective coupling constants.
The temperature expansion also clearly exhibits the limitations of
the method. The truncated series can be trusted only at low temperatures,
where the first term represents the dominant contribution. According to the
above formula, the quark condensate drops to about half of the vacuum
expectation value when the temperature reaches
$160\;\mbox{MeV}$ --- the formula does not make much sense beyond this
point. In particular, the behaviour of
the quark condensate in the vicinity of the chiral phase transition is
beyond the reach of the effective theory discussed here.
\section{Nonrelativistic effective Lagrangians}
\label{nr}
The fact that symmetries may break down spontaneously was discovered
in condensed matter physics. Also,
the phenomena associated with the propagation of sound were among the
very first to be analyzed in terms of an effective field theory.
The main difference to the
situation in particle physics is that the ground state, which forms, when the
number of electrons and baryons is fixed at a nonzero value, fails to be
Lorentz invariant: The rest frame singles out a preferred frame of reference.
The Hamiltonian is invariant under the Poincar\'{e} group G, but the
ground state is invariant only under a subgroup thereof,
$\mbox{H}_\im{solid}\subset\mbox{H}_\im{fluid}=
\mbox{H}_\im{gas}\subset \mbox{G}$.
As is well-known, the {\it phonons} may be viewed as Goldstone bosons
generated by this spontaneous symmetry breakdown. Their properties are rather
special, however, because they originate in a space-time symmetry rather
than an internal one:  The corresponding conserved "currents" are the
components of the energy-momentum tensor $\theta^{\mu\nu}$ and their number
is smaller than the dimension of the coset space G/H. I do not elaborate on
this further here, but refer to [7--9]. Instead, I add a
few remarks concerning
nonrelativistic {\it internal} symmetries, emphasizing the comparison with the
relativistic situation.

As an example, I consider the Heisenberg model, where
the dynamical variables form a lattice of spin operators
$\vec{s}_\im{n}$. The Hamiltonian of the model reads
\bdm  H=g\sum_\im{mn}
\vec{s}_\im{m}\cdot\vec{s}_\im{n}\co\edm where the sum runs over nearest
neighbours.
It is invariant under rotations of the spin directions, generated
by
\bdm \Qvec=\sum_\im{n}\vec{s}_\im{n}\fs\edm
Note that the corresponding group G = O(3) represents an internal
symmetry, because the space
lattice remains put. If the coupling constant $g$ is positive,
the interaction favours an antiparallel alignment of the spins, such that the
model shows the behaviour of an antiferromagnet. For positive coupling, the
ground state instead forms a configuration of parallel spins, like for
a ferromagnet. In either case, the ground state singles out a
direction and thus spontaneously breaks
the symmetry group of the Hamiltonian to the subgroup H = O(2) of the rotations
around this direction.
In the present case, the Goldstone bosons generated by the spontaneous
symmetry breakdown represent spin waves or magnons. The coset
space G/H is the unit sphere, such that the effective field is a unit
vector $\vec{U}(x)$ and carries two degrees of freedom.

The low energy behaviour of the model may again be analyzed in terms of an
effective Lagrangian. Consider first the antiferromagnetic case, $g\!>\!0$,
where the relevant
order parameter is the staggered magnetization.
For a cubic lattice, the leading terms in the derivative expansion of the
corresponding effective Lagrangian are given by
\bdm {\cal L}_\eff^2=\mbox{$\frac{1}{2}$}F^2\{\partial_t\vec{U}\!\!\cdot\!
\partial_t\vec{U}
-c^2\mbox{\raisebox{0.15em}{$\scriptstyle\sum$}}\hspace{-0.7em}
\mbox{\raisebox{-0.7em}{$\scriptstyle s$}}\;\,
\partial_s\vec{U}\!\!\cdot\!\partial_s\vec{U}\}\co\edm
where the sum extends over the three space directions.
The reflection symmetries of the lattice imply that the expression
is invariant
under space rotations. The constant $c$ is the spin wave velocity, while $F$
is related to the helicity modulus.
Evidently, the effective
Lagrangian is very similar to the one occurring in QCD or in the Higgs model.
A suitable change in scale takes the constant $c$ into the
velocity of light: The leading terms in the derivative expansion of
the effective Lagrangian relevant for an antiferromagnet are invariant
with
respect to Lorentz transformations. There is a difference in the structure
of the symmetry groups, as we are now dealing with the spontaneous breakdown
$\mbox{O(3)}\rightarrow\mbox{O(2)}$ rather than
$\mbox{O(4)}\rightarrow \mbox{O(3)}$.
There are two Goldstone bosons instead of three as in QCD.
Apart from that, however, the effective Lagrangians are the same.
As a consequence, the formula (\ref{cond1})
also holds for the staggered magnetization of an antiferromagnet, except that
the Clebsch-Gordan coefficients, which accompany the various powers of $T$ are
different, because the symmetry groups are not the same. The temperature scale
of the melting process is now set by the helicity modulus and is more than
eight orders of magnitude smaller than in the case of the quark condensate.
Otherwise, the behaviour of the two systems at low temperatures is
essentially the same.

Remarkably, the behaviour of a ferromagnet at low energies is quite
different.
Although the Hamiltonian differs from the preceding case only in the sign of
the coupling constant $g$, the correponding effective Lagrangian is not the
same. The groups involved in the spontaneous symmetry
breakdown are identical, such that
the Goldstone bosons are again described by a vector
field $\vec{U}(x)$ of unit length.
The difference in the low energy behaviour arises from the fact that for the
antiferromagnet, the mean value $\langle
\Qvec\rangle$ of the sum over all spins vanishes, while for the ferromagnet,
this is not the case:
The generators of the symmetry group give rise to an order
parameter [\hspace{0.05em}8--11].

In the relativistic domain, this cannot happen. The charges
of an internal symmetry are integrals over the time components of the
corresponding currents. For a Lorentz invariant
ground state, currents cannot pick up an expectation value. In the case of
a ferromagnet, however, the expectation values of the charges represent
the most important order parameter, the spontaneous magnetization,
\bdm \langle \Qvec\rangle=V\langle \Mvec\rangle\fs\edm
In this perspective, the antiferromagnet is exceptional: The
symmetry does not prevent the generators from picking up an expectation
value, but it does not ensure that this happens.
For an antiferromagnet, the quantity $\langle\Qvec\rangle$
happens to vanish, for dynamical reasons.

In the effective Lagrangian, the order parameter $\langle\Qvec\rangle$
manifests itself through a topological term, related to the Brower degree.
Like the Wess-Zumino term, this
contribution
is invariant under the symmetry group only up to a total derivative. While the
Wess-Zumino term only shows up at higher orders of the low energy expansion,
the one relevant for a ferromagnet contributes at leading order and thus
profoundly
modifies the low energy structure of the system.
Although the number of effective {\it fields} is the same as in the case of the
antiferromagnet, the number of Goldstone {\it particles} is different and the
dispersion laws are not the same, either:
antiferromagnetic magnons possess two polarization states
and the dispersion is of the
form $\omega(\vec{k})\!\propto\!|\vec{k}|$,
while for a ferromagnet, only one polarization occurs and
$\omega(\vec{k})\!\propto\!|\vec{k}|^2$. In a sense, the
difference in the low energy
structure of a ferromagnet and an antiferromagnet is more pronounced than
the one between an antiferromagnet and QCD.
\section{Concluding remarks}
\label{co}
Spontaneously broken symmetries play an important role, in condensed matter as
well as in particle physics. The low energy properties of the Goldstone bosons
generated by the symmetry breakdown may be worked out by means of the effective
field theory methods, invented in the 1960's. Since then,
the effective Lagrangian technique has been developed into an efficient and
mathematically precise tool, used extensively, e.g., to analyze the low energy
structure of QCD. Several applications
to the strong,
electromagnetic and weak interactions of the pseudoscalar mesons were worked
out in detail. In particular,
rare decays and anomaly driven processes provide sensitive tests of the theory.
In addition, the thermal properties of the hadronic phase \cite{Gerber}, the
mass generating
sector of the Standard Model \cite{mass generation} and finite size effects in
models with a spontaneously broken symmetry  \cite{finite size}
have been analyzed with this
method. Much remains to be done in this field, however, also in view of the
low energy precision experiments planned at various laboratories. Once lattice
simulations of QCD reach the domain, where the long range phenomena associated
with the spontaneous breakdown of chiral symmetry
become visible, the method should also prove to be an efficient tool to
account for the corresponding finite size effects.

In condensed matter physics, spontaneous breakdown occurs for internal as well
as space-time symmetries. In the language of the relevant effective
Lagrangian, the
good old description of the behaviour at long wavelength corresponds
to the leading term of the derivative expansion.
In the case of the antiferromagnet, the effective Lagrangian proved to
be very useful also beyond leading order. For other nonrelativistic systems,
such as ferromagnets, the higher order terms, due to the quantum fluctuations
of the effective field, yet need to be worked out.
Nonrelativistic kinematics is
less restrictive than Lorentz invariance and allows the generators of the
symmetry to become order parameters. In the effective Lagrangian, these are
represented by a term of topological nature, which does not occur in particle
physics.

The method has its limitations. In particular, it is useful only at low
momenta, small quark masses, weak external magnetic fields, low
temperatures and large volumes. The behaviour of
the quark condensate in the vicinity of the chiral phase transition, e.g., is
beyond the reach of this technique. Another limitation arises from the fact
that the quantum
fluctuations of the effective field play an important role in the
systematic low energy analysis. These can only be worked out if the dynamics of
the effective degrees of freedom may be formulated in terms of a Lagrangian.
A phenomenological description of the dissipative
effects
generated by friction is beyond this framework, because frictional
forces cannot be accounted for in terms of a Lagrangian.
\section*{Acknowledgement}
It is a pleasure to thank Haridas Banerjee, Pasha Kabir and especially Samir
Mallik for the warm hospitality during our visit to this fascinating
country.

\end{document}